\documentclass[twocolumn,aps,floats,superscriptaddress,showpacs]{revtex4}
\usepackage{amsmath}

\usepackage{epsfig}
\usepackage{graphicx}
\usepackage{dcolumn}
\usepackage{bm}

\def\gapp{\lower.35em\hbox{$\stackrel{\textstyle>}{\sim}$}}
\def\lapp{\lower.35em\hbox{$\stackrel{\textstyle<}{\sim}$}}

\begin{document}
\bibliographystyle{apsrev}
%

\newcommand{\beq}{\begin{equation}}
\newcommand{\eeq}{\end{equation}}
\newcommand{\beqa}{\begin{eqnarray}}
\newcommand{\eeqa}{\end{eqnarray}}
\newcommand{\da}{^\dagger}
\newcommand{\wh}{\widehat}

\newcommand{\Om}{\Omega}
\newcommand{\om}{\omega}
\newcommand{\tr}{{\rm tr}}
\newcommand{\intf}{\int_{-\infty}^\infty}
\newcommand{\into}{\int_0^\infty}
\newcommand{\I}{{\mathcal I}}
\newcommand{\G}{{\mathcal G}}
\newcommand{\A}{{\mathcal A}}
\newcommand{\F}{{\mathcal F}}
\newcommand{\C}{{\mathcal C}}
\newcommand{\x}{\vec x}
\newcommand{\X}{\vec X}
\newcommand{\q}{\vec q}

\renewcommand{\=}{\!\!=\!\!}

\def\simleq{\; \raise0.3ex\hbox{$<$\kern-0.75em
      \raise-1.1ex\hbox{$\sim$}}\; }
\def\simgeq{\; \raise0.3ex\hbox{$>$\kern-0.75em
      \raise-1.1ex\hbox{$\sim$}}\; }

\def\la{{\langle}}
\def\ra{{\rangle}} \def\vep{{\varepsilon}} \def\y{\'\i}
\def\half{{1\over 2}}
\def\an{|\Phi _N \rangle }
\def\bn{|\Phi ' _{N'}\rangle }
\def\na{ \langle \Phi _N|}
\def\nb{ \langle \Phi'_{N'} |}

\def\ov{\over}
\def\non{\nonumber }
\def\beq{\begin{equation} }
\def\eeq{\end{equation} }
\def\beqa{\begin{eqnarray}}
\def\eeqa{\end{eqnarray}}
\def\del{\partial }
\def\D{\Delta}
\def\a{\alpha }
\def\am{\alpha^\mu }
\def\Xm{X^\mu}
\def\Xn{X^\nu}
\def\d{\textrm{d}}
\def\b{\beta}
\def\t{\tau}
\def\e{\epsilon}
\def\g{\gamma}
\def\s{\sigma}
\def\med{\frac{1}{2}}
\def\go{\vec g_1}
\def\gt{\vec g_2}
\def\eq{\!\! =\!\!}

\title{Existence and topological stability of Fermi points in
multilayered graphene}
\author{J. L. Ma\~nes}
\affiliation{Departamento de F\'{\i}sica  de la Materia Condensada\\
Universidad del Pa\'{\i}s Vasco,
Apdo. 644, E-48080 Bilbao, Spain}
\author{F. Guinea}
\affiliation{Unidad Asociada CSIC-UC3M,
Instituto de Ciencia de Materiales de Madrid,\\
CSIC, Cantoblanco, E-28049 Madrid, Spain.}
\author{ Mar\'{\i}a A. H. Vozmediano}
\affiliation{Unidad Asociada CSIC-UC3M,
Universidad Carlos III de Madrid, E-28911
Legan\'es, Madrid, Spain. }

\date{\today}
\begin{abstract}
We study the existence and topological stability of Fermi points
in a graphene layer and stacks with many layers.
We show that the discrete symmetries
(spacetime inversion) stabilize the
Fermi points in monolayer, bilayer and multilayer
graphene with orthorhombic  stacking. The bands near $k=0$ and $\epsilon=0$
in multilayers with the Bernal stacking depend on the parity of the number of
layers, and Fermi points are unstable when the number of layers is
odd. The low energy changes in the
electronic structure induced by commensurate perturbations which mix the two
Dirac points are also investigated.
\end{abstract}
%
\pacs{75.10.Jm, 75.10.Lp, 75.30.Ds}

\maketitle
\section{Introduction.}

The recent synthesis of monolayer graphite~\cite{Netal04,Netal05b}
(graphene),  the experimental ability to manipulate few layer
samples~\cite{Betal04,Netal05,Zetal05} and the observations of
quasi two dimensional behavior in graphite~\cite{Ketal03}, have
awaken an enormous interest in these systems. The conduction band
of graphene is well described by a tight binding model which
includes the $\pi$ orbitals which are perpendicular to the plane
at each C atom\cite{W47,SW58}. This model describes a semimetal,
with zero density of states at the Fermi energy, and where the
Fermi surface is reduced to two inequivalent K-points located at
the corners of the hexagonal Brillouin Zone. The low-energy
excitations with momenta in the vicinity of any of the Fermi
points have a linear dispersion and can be described by a
continuum model which reduces to the Dirac equation in two
dimensions~\cite{GGV92,GGV93},  what has been tested by recent
experiments~\cite{Netal05,Zetal05,Zetal06}. Fermi points have also
been found in the modelling of the low energy band structure of
multilayer systems both theoretically~\cite{NCPG06,NCGP06} and in
experiments~\cite{Netal06b,Zetal06b}. A crucial issue for both
theory and technology is the possibility of controlling the
opening of a gap in the samples. From a theoretical point of view,
the gap is related to the chiral symmetry breaking and mass
generation, a classical -unresolved- problem that has been
explored at length in planar QED~\cite{AN88,Khv01}. For the
applications it is by now clear that opening a gap in monolayer
graphene will be a difficult task and efforts are concentrated on
multilayer structures~\cite{NCGP06,Oetal06,Cetal06}.

In this paper we analyze the stability of the Fermi points under small perturbations
using very basic topological concepts~\cite{Naka}. We find that the Fermi points are protected
by the discrete symmetries (translational invariance and space and time inversion)
in the monolayer, bilayer AB and multilayers with $ABCA \cdots$ (rhombohedral)
stacking. The stability of Fermi points in stacks with the Bernal stacking,
$ABAB \cdots$, depends on the parity of the number of layers. We also discuss
the changes in the low energy and low momenta properties induced by
commensurate perturbations which hybridize the two $K$ points and partially
break translation invariance. We will not analyze here in detail the effects
of spin-orbit coupling, which my lead to additional changes at low temperatures\cite{KM05,HGB06,Metal06b,Yetal07}.

The analysis reported here will  be useful for the construction of
continuum theories
for long wavelength spatial perturbations, and for the study of the degeneracy
and spectrum of the low index Landau levels in a magnetic field.

\section{ Electronic structure and stability in graphene.} The
Fermi surface (FS) is a central concept in condensed matter that
controls the low-energy physics of the systems. In a Landau Fermi
liquid at T = 0, J. Luttinger \cite{L60} defined the FS of an
interacting Fermi system in terms of the single-particle Green's
function $G({\vec k}, \omega)$, as the solution of the equation
$G^{-1}({\vec k}, 0) = 0 $ and showed that it encloses the same
volume, equal to the fermion density $n$, as in the noninteracting
system. The robustness of the Fermi liquid idea has been
understood recently in the context of the renormalization group
where the Fermi and Luttinger liquids are seen as infrared fixed
points\cite{P93,S94}. In  recent works\cite{V03,V06} Volovik has
emphasized the idea of the topological stability of the Fermi
surface as the origin of the robustness of the Fermi liquid and
has suggested a classification of general fermionic systems in
universality classes dictated by momentum space topology. A more
recent proposal relates the stability of Fermi surfaces with
K-theory, a tool used to classify D-brane charges in string
theory\cite{H05}. The idea behind the topological stability is to
study the zeroes of the matrix $G^{-1}_0({\vec k}, \omega)$ (free
inverse propagator) that can not be lifted by small perturbations.
Here we will analyze the stability of the Fermi points of single
and multilayer graphene, where  the discrete symmetries of the
system play a principal role. Although we will restrict ourselves
to perturbations that can be studied within the context of a
single-particle effective hamiltonian, the extension to
self-energy induced  perturbations is rather straightforward and
will be reported elsewhere\cite{us07}.

\begin{figure}[!t]
\begin{center}
\includegraphics[width=5cm]{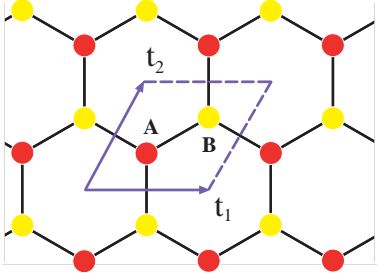}
\caption[fig]{\label{DL}(Color online) Direct lattice and unit
cell for monolayer graphene. }
\end{center}
\end{figure}
\begin{figure}[!t]
\begin{center}
\includegraphics[width=4cm]{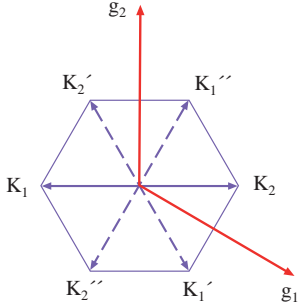}
\caption[fig]{\label{BZ}  (Color online) First Brillouin zone and
Fermi points. The vectors $\vec K\,'_1,\vec K\,''_1$ ($\vec
K\,'_2,\vec K\,''_2$) are equivalent  to $\vec K_1$ ($\vec K_2$).}
\end{center}
\end{figure}

As shown in Fig. \ref{DL}, monolayer graphene consists of a planar honeycomb  lattice of
carbon atoms. Corresponding to the two atoms in the unit cell, one
may define two Bloch wave functions to be used in a variational
(tight-binding) computation of the spectrum \beq\label{bloch}
\Phi_i (\vec K)=\sum_{\vec t} e^{i \vec K\cdot(\vec r_i+ \vec t)}
\Phi (\vec r-\vec r_i-\vec t) \; \;\; , \; \;\; i=A,B \eeq where
the sum runs over all the points in the direct lattice, i.e.,
$\vec t = n_1 \vec t_1+n_2 \vec t_2$, $(\vec r_A,\vec r_B)$ are
the positions of the atoms in the unit cell,  and $\Phi(\vec r)$
is a real ($\pi$-type) atomic   orbital. As is well known, a
simple tight-binding computation~\cite{W47,SW58} yields  a
spectrum with two Fermi points  located at $\vec K_1 = -2\go/3-
\gt/3$ and $\vec K_2= -\vec K_1$, where $\vec t_i\cdot\vec g_j=2\pi \delta_{ij}$ (see Fig. \ref{BZ}). Near the
two Fermi points, the hamiltonian can be linearized, and using
appropriate units one finds \beq\label{lin} H(\vec K_1+\vec
k)\sim\left(
\begin{array}{cc} 0 & k^*
\\ k & 0 \end{array} \right)=k_x\sigma_x+k_y\sigma_y ,
\eeq
and $H(-\vec K_1+\vec k)\sim - k_x \sigma_x + k_y \sigma_y$.
where $k\equiv k_x+ik_y$ and $\sigma_i$ are the Pauli matrices.
Thus, the low energy electronic excitations behave like massless
Dirac fermions with relativistic spectrum $E=\pm |k|$.

Under a $k$-independent, translationally invariant  perturbation
 \beqa\label{const} & H(\vec
K_1+\vec k)\to\left( \begin{array}{cc} a_z & k^*+a^*
\\ k+a & -a_z \end{array} \right)
\eeqa
where $a\equiv a_x+i a_y$,  the spectrum becomes
$E=\pm\sqrt{a_z^2+|k+a|^2}$ and a gap $2| a_z|$is generated.
This is consistent with Horava's general arguments~\cite{H05},
which show that Fermi points for Dirac fermions in two dimensions are unstable.
However, this is not the end of the story, since the
existence of discrete symmetries can sometimes stabilize
the Fermi loci. In the case of the graphene, this role
is played by time-reversal $T:t\to -t$ and spatial inversion
$I:(x,y)\to(-x,-y)$. The reality of the $\pi$~orbitals implies that
 time reversal  merely reverses $\vec K$
\beq\label{compl}
T\Phi_i(\vec K) =\Phi^*_i(\vec K)=\Phi_i (-\vec K)
\eeq
whereas the
spatial inversion  also exchanges the two types of
atoms
\beq\label{I}
I\Phi_A (\vec K)=\Phi_B (-\vec K)\;\;\; , \;\;\; I\Phi_B (\vec K)=\Phi_A (-\vec K)
\eeq
Invariance under these symmetries imposes the following constraints on the hamiltonian
\beqa\label{trev}
T:\;  H (\vec K)&=&H^* (-\vec K)\non\\
I:\;H (\vec K)&=& \sigma_x H(-\vec K) \sigma_x
\eeqa
Although these are useful properties that relate the hamiltonians at opposite values of $\vec K$,
what we need is a constraint on the form of $H(\vec K)$.
This is obtained by combining time reversal with the
spatial inversion
\beq\label{TI}
TI:\; H(\vec K)=\sigma_x H^*(\vec K) \sigma_x
\eeq
implying $H_{11}(\vec K)=H_{22}(\vec K)$. This enforces
  $a_z\!=\!0$ in~(\ref{const}) and  we see that no gap opens if
   the perturbation preserves the space-time inversion $TI$.

  This has an interesting topological interpretation, which extends
  the previous arguments to $k$-dependent ---but translationally invariant---  perturbations.
  The low energy hamiltonian $H(\vec K_1+\vec k)$ in~(\ref{lin})
  defines a  map from the circle $k_x^2+k_y^2=R^2$ to the space
  of $2\times 2$ hamiltonians
 $H=\vec h \cdot\vec \sigma$:
 \beq\label{map1}
 k=R e^{i\theta} \to (h_x,h_y, h_z) =R (\cos\theta ,\sin\theta,0)
 \eeq
  Since Fermi points correspond to zeroes of the determinant
$ -Det(H)=h_x^2+h_y^2+h_z^2$,
  a perturbation will be able to create  a gap only if the loop
  represented by the map~(\ref{map1}) is contractible in the space
  hamiltonians with non-vanishing determinants, which 
  is just $R^3-\{ 0\}$. This is clearly the case, since
$ \pi_1(R^3-\{ 0\})=\pi_1(S^2)=0$.
On the other hand, hamiltonians invariant under $TI$ are represented
 by points in $R^2$, and   we have
 \beq\label{hom2}
 \pi_1(R^2-\{ 0\})=\pi_1(S^1)=Z
 \eeq
This means that non-trivial maps such as the ones implied
by~(\ref{lin}) can only be extended to the interior of the
circle by going through the origin, i.e., by having at
least one zero. This precludes the creation of a gap.

 Note that the maps defined by the low energy hamiltonian
 in the proximity of the two Fermi points
 \beq
  H(\pm\vec K_1+\vec k)\; :  \; k=R e^{i\theta} \to h_x+ih_y=\pm R e^{\pm i\theta} 
  \eeq
have opposite winding numbers $N=\pm 1$, which can  be computed by
the formula \beq\label{wind} N={1\over 4\pi i}\int_0^{2\pi} d
\theta\, \mathrm{Tr} (\sigma_z H^{-1} \partial_\theta H) \eeq

The fact that the two Dirac points carry opposite charges
suggests that they could annihilate mutually if brought
together by a perturbation. Any external potential commensurate with
 the honeycomb lattice, which has a finite Fourier component at the
 wavevector $\vec{G} = \vec{K}_1 - \vec{K}_2$, induces terms which hybridize
 the two Dirac points and it will lead to the possibility of a gap.
 We can
 compute all the possible perturbations which are compatible with the
 symmetries of the lattice. The most general $( 4 \times 4 )$ hamiltonian
 including perturbations at $\vec{G} = 0$ and  $\vec{G} = \vec{K}_1 -
 \vec{K}_2\equiv -\vec K_1$ is:
\begin{equation}
{\cal H} \equiv \left( \begin{array}{cccc} 0& k^* + Q^*_1&Q_2
    &Q_4 \\ k + Q_1 &0 &Q_4 &Q_3 \\ Q_2^* &Q_4^* &0 &-
    k+Q_1 \\ Q_4^* &Q_3^* &- k^*+Q_1^*
  &0 \end{array} \right)
  \label{Hpert}
\end{equation}
where $Q_1=Q_1^x+i Q_1^y$ transforms according  to the
  $E_2$ representation of the $C_{6v}$ symmetry group\cite{note}
at the \hbox{$\Gamma ( \vec{K}
  = 0)$} point. $Q_2$ and $Q_3$ belong to the $E$ representation of the
  $C_{3v}$ group at $\vec{K}_1$, and $Q_4$ to the $A_1$ representation of the
  same group (see~\cite{BC72} for notation).
At this point it is worth noticing a point on notation. When
grouping the  hamiltonians attached to the two Dirac points
($K_{1,2}$) into a 4-dimensional matrix it is a common practice to
reverse the order of the sublattices (A,B) in one of the Fermi
points in such a way that the 4-dimensional wavefunctions have
the form
$$\psi=(\Phi_{K_1,A},\Phi_{K_1,B},\Phi_{K_2,B},\Phi_{K_2,A}).$$
If this is done the topological structure of the hamiltonian is
messed up and the computation of charges becomes  less clear. For this reason,
we follow instead the convention in~\cite{KM05}, where
$$\psi=(\Phi_{K_1,A},\Phi_{K_1,B},\Phi_{K_2,A},\Phi_{K_2,B}).$$
This is also important if one tries to compare the analysis of the
perturbations written in eq. (\ref{Hpert}) with the ones produced
by the different types of disorder \cite{A06,Mcetal06,AE06}.

\begin{figure}[!t]
\begin{center}
\includegraphics[width=5cm]{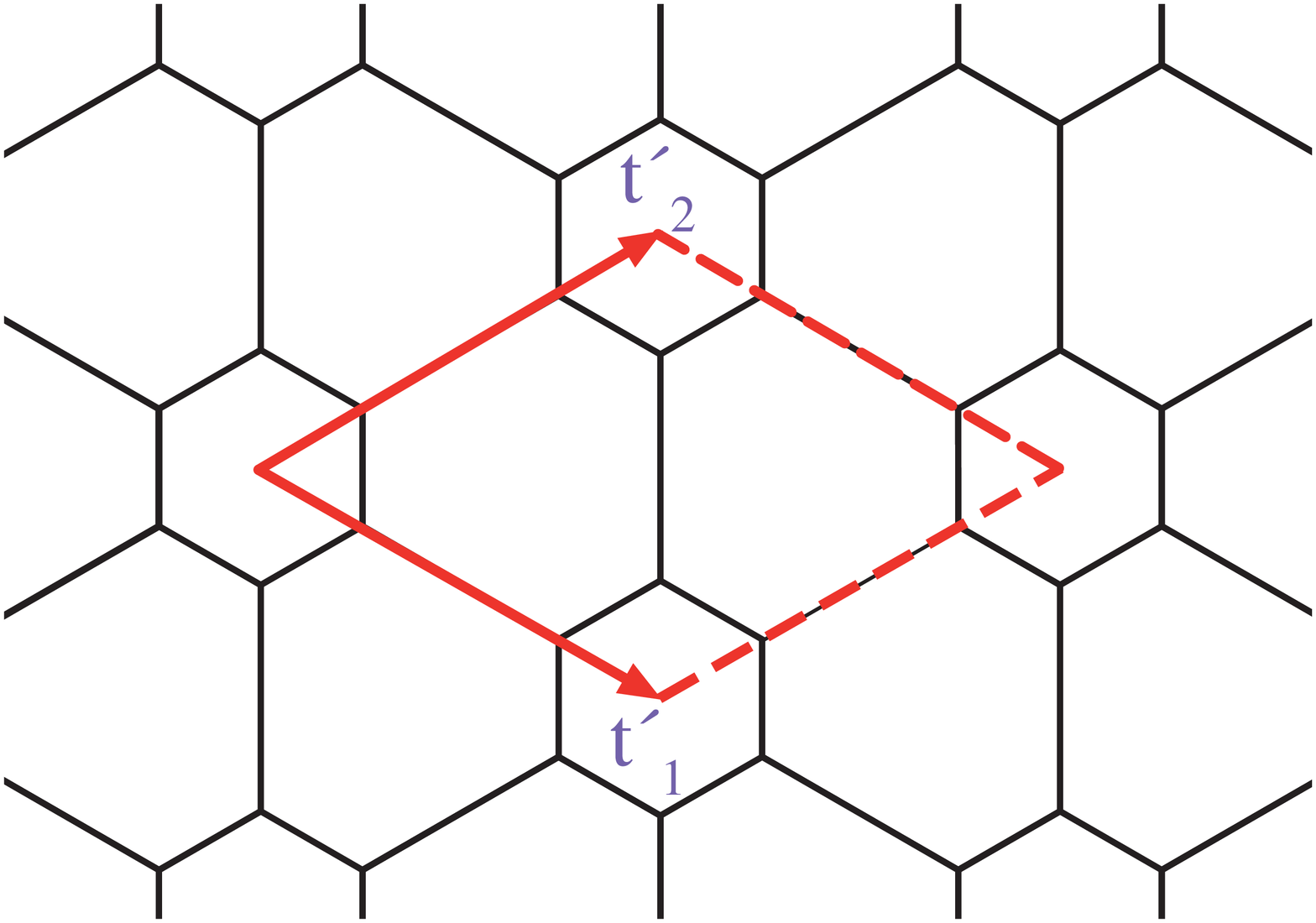}\\
\caption[fig]{\label{f2} (Color online)Distortion caused by the
condensation of $A_1$ mode with real $Q_4$ and new unit cell.}
\end{center}
\end{figure}

  The perturbation given by $Q_1^x$ and $Q_1^y$ shifts the Dirac
  points, but  does not open a gap. In fact, the only parameter which opens a gap is~$Q_4$. When only
  $Q_4$ is different from zero, the spectrum  becomes $E=\pm\sqrt{|Q_4|^2+|k|^2}$ and,
  for $Q_4$ real, the deformation of the lattice is given in Fig.~\ref{f2},
where one can see that no point symmetry is broken.
This shows, in particular, that invariance under space-time
inversion  is \textit{not} enough to guarantee the stability of
Fermi points ---translation invariace plays a crucial role: $TI$
by itself makes de Fermi points \textit{individually} stable, but
they may still annihilate against each other in the presence of a
perturbation that breaks translation invariance. A distortion of
the type of $Q_4$ can be induced by a substrate with a periodicity
commensurate with the lattice, or by the effect of one layer on
another when there is a lattice mismatch between them, as in
samples grown on a substrate~\cite{Betal04,Betal06}. It can be
responsible for the gap observed recently photoemission
experiments\cite{L07}. It is worth noting that the perturbation
denoted here $Q_4$ has been studied in a graphene ribbon
in~\cite{GLV06}.

\begin{figure}[!t]
\begin{center}
\includegraphics[width=4cm]{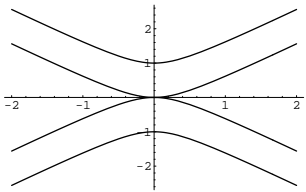}
\caption[fig]{\label{Q3}  Energy bands along the line $(0,k_y)$
for $Q_3=1$, $Q_4=0$.}
\end{center}
\end{figure}

When only $Q_2$ or $Q_3$ are
  different from zero, we find:
\begin{equation}\label{bila}
\epsilon_k =\pm \frac{|Q_{2,3}|}{2} \pm
\sqrt{\frac{|Q_{2,3}|^2}{4} + | \vec{k} |^2}.
\end{equation}
The energy bands are represented in Fig. \ref{Q3} for the
particular case $Q_2=0, Q_3=1$. We can see that  the spectrum is
the same as the one obtained in a simple model for a bilayer
system~\cite{MF06}, which will be discussed later.  A complete analysis of the most general
perturbation of the form~(\ref{Hpert})   will be given elsewhere~\cite{us07}.

In the absence of time reversal symmetry, other perturbations are possible, such as
\begin{equation}
\delta {\cal H} = B_1 \sigma_z  \tau_z + B_2 \sigma_y \tau_y
\label{mag_field}
\end{equation}
where $\sigma$ and $\tau$ are Pauli matrices whose entries are the sublattice
and K point indices respectively, and $B_1$ and $B_2$ transform like the $z$ component of a magnetic field
 and are odd under time inversion. Note that the
first term is the orbital part of the intrinsic spin orbit coupling in
graphene~\cite{KM05,HGB06,Metal06b,Yetal07}, and it opens a gap. The second term should appear in a
general spin orbit hamiltonian which takes into account the coupling between
the two $K$ points.

\begin{figure}[!t]
\begin{center}
\includegraphics[width=4cm]{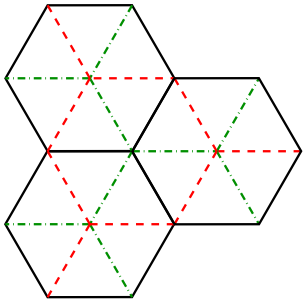}
\caption[fig]{\label{stacking} (Color online) Skecth of the three
possible positions of a given
  layer with respect to the others in a graphene stack. Bernal stacking
  ($1,2,1,2, \cdots$, is described by two inequivalent planes, while
  orthorhombic stacking, $1,2,3,1,2,3 \cdots$, requires the three
  inequivalent ones.}
\end{center}
\end{figure}

\section{ Electronic structure and stability in multilayered
graphene.} The case of the multilayer is more interesting. We will
concentrate on the ability of the $TI$ invariance to prevent the
creation of a gap. For the sake of definiteness, only staggered
($ABA$) and rhombohedral ($ABC$) stacking will be considered. The relative
orientations of the $ABC$ planes are sketched in Fig.[\ref{stacking}]. The
two inequivalent atoms in layer $n$ will be denoted ($A_n,B_n$).
Our conventions are such that the couplings of an $A_n$ ($B_n$)
atom to the  three in-plane nearest neighbors are shaped as a Y
(inverted Y), independently of $n$. Thus, the low energy limit of
the ``free'' hamiltonian obtained by neglecting inter-layer
couplings is block-diagonal, with $2\times 2$ blocks given
by~(\ref{lin}).

The simplest model introduces interlayer hoppings $t$ only between
nearest neighbors. The resulting hamiltonian for bilayer graphene in the vicinity of the
$K_1$ Fermi point is
\beq\label{bil}
\mathcal{H}(k)=\left(
\begin{array}{llll}
 0 & k^* & 0 & t \\
 k & 0 & 0 & 0 \\
 0 & 0 & 0 & k^* \\
 t & 0 & k & 0
\end{array}
\right)
\eeq
and the energy bands are
given by~(\ref{bila}) with the replacement $|Q_{2,3}|\to t$. In the limit $E\ll t$, one can
obtain an effective hamiltonian~\cite{MF06} for the lowest energy
bands.
To this end, reorder the wavefunctions according $(A_1,B_1,A_2,B_2)\to (A_2,B_1,A_1,B_2)$,
so that in the new basis the hamiltonian becomes
\beq
\mathcal{H}(k)=\left(
\begin{array}{llll}
 0 & 0 & 0 & k^* \\
 0 & 0 & k & 0 \\
0 & k^* & 0 & t \\
 k & 0 & t & 0
\end{array}\right)\equiv\left(
\begin{array}{ll} H_{11} &H_{12}\\
 H_{21} &H_{22}
 \end{array}
\right)
\eeq
where $H_{ij}$ is a $2\times 2$ block.
The  identity
\begin{eqnarray}
& Det(\mathcal{H}-E)\\ \nonumber =
&Det\Bigl(H_{11}-H_{12}(H_{22}-E)^{-1}H_{21}-E\Bigr)\,Det(H_{22}-E)
\end{eqnarray}
shows that, for $E\ll t$, the substitution $H_{22}-E\to
H_{22}$ reduces the computation of the lowest energy bands to the
diagonalization of the $2\times 2$ effective hamiltonian
 \beq
 \label{bieff}
 \mathcal{H}^{eff}\equiv H_{11}-H_{12}H_{22}^{-1}H_{21}=-{1 \over t}\left(
\begin{array}{cc}
0 & k^{*2}\\
k^{2} & 0
 \end{array}
\right)
\eeq
This effective hamiltonian involves only the atoms $(A_2,B_1)$,
which are not linked by $t$  and give rise to  bands with zero energy
at the Fermi points. Since $(A_2,B_1)$ are interchanged under spatial
inversion \hbox{$\vec r\to -\vec r$},  the combined $TI$-invariance imposes a constraint
\hbox{$\mathcal{H}^{eff}(k)=\sigma_x \mathcal{H}^{eff*}(k)\sigma_x$}
identical  to~(\ref{TI}).  This implies
$\mathcal{H}_{11}^{eff}(k)=\mathcal{H}_{22}^{eff}(k)$,
 which shows that no gap can open.
 According to~(\ref{wind}) the  topological charge  for
 the $\vec K_1$ Fermi point is $+2$  and, by time reversal
 invariance, the charge for
  $-\vec K_1$ is $-2$. Thus, as in the case of monolayer graphene,
  the Fermi points are stable under perturbations that preserve $TI$ and translation invariance. For
  instance, a
  perturbation like trigonal warping~\cite{MF06} changes the off diagonal
  elements in eq.(\ref{bieff}), $k^2 \rightarrow k^2 + v_3 k^*$ and splits the
  Fermi point of charge $Q=+2$ into three Dirac points away from
  the $K$ point, and charge $Q=+1$, and another Dirac point at the $K$ point
  and $Q=-1$, but the total charge is conserved and no gap opens. However, a perturbation hybridizing
  $\vec K_1$ and $-\vec K_1$ or one breaking $TI$ might lead to a gapped system with no Fermi  points at all.
  A physical example is provided by the experiment described in
  \cite{Oetal06} where a gap is controlled by changing the carrier
  concentration in each layer.

  This analysis can be easily generalized to multilayer graphene with
  rhombohedral stacking. This type of staking includes the links
  $(B_1-A_2, B_2-A_3,\ldots, B_{N-1}-A_N)$  and the effective hamiltonian,
  which  involves  only the unlinked  atoms $(A_1,B_N)$, is given by
 \beq\label{muleff}
\mathcal{H}^{eff}\=-{1 \over t^{N-1}}\left(
\begin{array}{cc}
0 & k^{*N}\\
k^{N} & 0
 \end{array}
 \right)
\eeq
The topological charge for the $\vec K_1$ ($-\vec K_1$) Fermi point is $+N$ ($-N$).
As the point group for multilayer
graphene with rhombohedral stacking is $D_{3d}$, which contains
the inversion~$I$, the system is invariant under $TI$, which interchanges $(A_1,B_N)$,
and the whole argument goes through
as before. Thus we conclude that the Fermi points for
multilayer graphene with rhombohedral stacking are
stable against  perturbations that respect $TI$ and translation invariance.

The situation is very different for $ABA$ stacking.
An $N$-layer graphene stack is invariant under
the spatial inversion $I$ only for even $N$, where the point group is
$D_{3d}$, while it is $D_{3h}$ for odd $N$.

\begin{figure}[!t]
\begin{center}
\includegraphics[width=6cm,angle=-90]{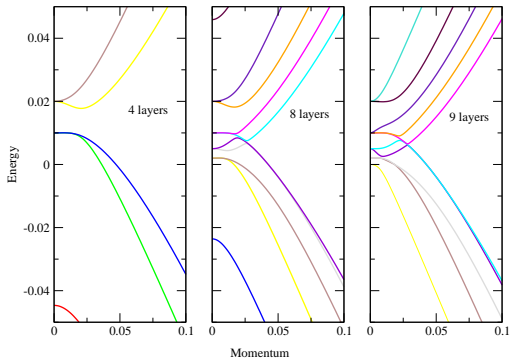}\\
\caption[fig]{\label{stack_4_8}(Color online) Energy bands close
to $\epsilon = 0$ for a stack with 4 layers (left), 8 layers
(center), and 9 layers (right). All stacks have the Bernal
stacking. The Fermi velocity is $v_F = 1 , t = 0.1$ and a layer
dependent shift has been included: $\epsilon_n \equiv \{ 0.02 ,
0.01 \}$ (left), $\epsilon_n \equiv \{ 0.02 , 0.01 , 0.005 , 0.002
\}$ (center), and $\epsilon_n \equiv \{ 0.02 , 0.01 , 0.005 ,
0.002 , 0\}$ (right).}
\end{center}
\end{figure}

The $2N$ eigenstates in a stack with $N$ layers at $k=0$
can be divided into two sets: $N$ states at the
orbitals connected by the interlayer hopping $t$, and $N$ states in the other
sublattice sites of each layer. If we only consider the hopping $t$, the
first $N$ states acquire a dispersion~\cite{GCP06}, lying in the range $- 2 t
\le \epsilon \le 2 t$. The other $N$ states are degenerate with $\epsilon = 0$. A
perturbation compatible with all the symmetries of the stack is a layer
dependent shift of the onsite energies. This shift can be arbitrary, except
for the twofold degeneracy related with the equivalence between layers which
are symmetrically placed around the center, $\epsilon_n = \epsilon_{N-n+1}$.
This is illustrated in Fig.[\ref{stack_4_8}].

The results in Fig.[\ref{stack_4_8}] show a gap at half filling
in the stack with four layers,
and overlapping bands at all energies for the stack with eight and nine
layers.
Note that the gap in the stack with four layers does not require the
existence of an external electric field,
which will break the equivalence of the layers at
opposite sides of the stack. In all stacks
with an even number of layers, the two bands which start  at the onsite energy of layers $n$ and
$N - n + 1$ are  degenerate at $k =0$.  The
effective $2 \times 2$ hamiltonian describing the bands near these degeneracy
points have off diagonal elements with a non trivial phase as in
eq.(\ref{bieff}). Hence, the degeneracy has topological charge $Q=2$, and it
cannot removed by perturbations compatible with the symmetries of the stack.
For even $n$, an explicit computation shows that these two bands
will have curvatures with the same sign near $k=0$. Thus, the corresponding
degeneracy points do not represent Fermi points. For odd numbered layers,
the two bands disperse in opposite directions away from $k=0$, and the
degeneracy points become stable Fermi points at the appropriate
doping.  Note that
the symmetries of the system allow for a direct trigonal-like coupling
between the two layers, which will split  this Fermi point and give rise
to four Fermi points showing linear dispersion, as in the bilayer. On
physical grounds, this coupling will be negligible, unless the two layers are
contiguous.

The low energy bands in a stack with an odd number of layers also
contain doubly degenerate states at $k=0$, associated to the equivalence
between layers at opposite sides of the stack. But   in this case the inversion $I$ is not part of
the symmetry group $D_{3h}$ of the stack, no invariance under $TI$ can be imposed and,
as a consequence,  the first homotopy group $\pi_1$ is trivial.
This means that no conserved topological charge exists.
 Hence, a gap
may open at $k=0$ when other perturbations consistent with the
symmetries of the stack are included. Concretely, a  direct coupling between orbitals in
the same sublattice in layers separated by an odd number of other layers will
open a gap. Such a coupling, between layers which are second nearest
neighbors, has been proposed in graphite~\cite{DSM77}.


\section{ Conclusions.}
We have presented a classification of the
bands at low momenta and low energy of graphene layers and stacks
with many layers.

Each Fermi point in single
layer graphene is stable against perturbations which preserve the discrete
$TI$ symmetry, and which do not mix the two Fermi points. A magnetic field,
for instance, induces a gap in the spectrum, see eq.~(\ref{mag_field}). This
term arises from the discreteness of the lattice, and it should be of higher
order than the minimal coupling which leads to the formation of Landau
levels. Combining this and dimensional arguments, we expect it to be $B_1
\propto v_F / l_B \times ( a / l_B )$, where $a$ is the lattice constant, and $l_B =
\sqrt{(e B ) / ( c \Phi_0 )}$ is the cyclotron radius. Thus, for $B \sim
10$T, we have $B_1 \approx 0.1$meV.

We have also classified the long wavelength perturbations commensurate with
the graphene lattice, which can hybridize the two $K$ points. Some of these
perturbations open a gap in the spectrum, while others shift the position of
the Dirac points. We expect that their strength will decay like a power law
with the wavelength of the distortion.

The low energy and low momentum spectrum of stacks with many graphene layers
depend on the stacking order and the number of layers. For the most common
case of the Bernal stacking, we find that layer dependent onsite energies
 lead to Fermi points with double degeneracy,
topological charge $Q= \pm 2$, and a parabolic dispersion in $k$. This
situation will be stable in stacks with a large (even)  number of layers.
In stacks with an odd number of layers, there is no conserved topological charge and  this
degeneracy will be broken by additional interactions. Stacks with
rhombohedral order lead to degenerate states with a large topological charge,
$Q=N$, which will give rise to the formation of a cascade of Fermi points
slightly away from $k=0$, with lower topological charges.

The most likely origin of the inequivalence between layers is the charge
accumulation at the layers close to the surface\cite{G06}. An induced doping of
$10^{10} - 10^{12}$ cm$^{-2}$ gives rise to shifts in the local potential of
$0.01 - 0.1$eV, so that the splittings associated to this effect can be
easily measurable. We find that a true gap opens, in the absence of an
external field which breaks spatial inversion,  only in a stack with four
layers and Bernal stacking. Finally, stacking defects, which break
the equivalence between pairs of layers,
will also break the degeneracy of all bands at $k=0$.

{\em Acknowledgments.}
It is a pleasure to thank J.J.~Manjar\y n and M.A.~Valle-Basagoiti
for  useful discussions. M. A. H. V. and F. G. are thankful to the MEC (Spain) for
financial support through grant  FIS2005-05478-C02-01
and the European Union Contract 12881 (NEST). J.L.M. has been
supported in part by the Spanish Science Ministry under Grant
FPA2005-04823. One of us (F. G.) also acknowledges support from the Comunidad
de Madrid, through the program CITECNOMIK, CM2006-S-0505-ESP-0337.

\bibliography{PRB}

\begin{thebibliography}{43}
\bibitem[{not()}]{note}
\bibinfo{note}{The actual symmetry group of monolayer graphene is $D_{6h}$.
  But, as long as we restrict ourselves to in-plane distorsions of the lattice,
  it is sufficient to consider the subgroup $C_{6v}$. That is what we do in
  this paper, since out-of-plane distorsions (phonons) can not couple linearly
  to electrons.}
\end{thebibliography}

\end{document}